\documentclass[12pt,preprint,graphicx]{aastex}

\slugcomment{Draft:  \today}

\lefthead{Smith, Tucker, Allam, \& Rodgers}
\righthead{CDF--S $u'g'r'i'z'$ Standards}
\begin{document}

\title{Local $u'g'r'i'z'$ Standard Stars in the Chandra Deep Field--South}

\author{J. Allyn Smith\altaffilmark{\ref{Wyoming},\ref{LANL},\ref{CTIO}},
Douglas L. Tucker\altaffilmark{\ref{Fermilab},\ref{CTIO}},
Sahar S. Allam\altaffilmark{\ref{NMSU},\ref{Fermilab}},
Christopher T. Rodgers\altaffilmark{\ref{Wyoming},\ref{CTIO}}
}
\altaffiltext{1}{Department of Physics \& Astronomy, P.O. Box 3905,
            University of Wyoming, Laramie, WY 82071 \label{Wyoming}}
\altaffiltext{2}{Current address: Mail Stop D-448; NIS-4,
            Los Alamos National Laboratory,
            Los Alamos, NM 87545 \label{LANL}}
\altaffiltext{3}{Visiting Astronomer, Cerro Tololo Inter-American Observatory.
            CTIO is operated by AURA, Inc. under contract to the
            National Science Foundation. \label{CTIO}}
\altaffiltext{4}{Fermi National Accelerator Laboratory, MS-127,
            P.O. Box 500, Batavia, IL 60510 \label{Fermilab}}
\altaffiltext{5}{Current address: Astronomy Department,
            New Mexico State University, Box 30001, Dept. 4500,
            1320 Frenger St., Las Cruces, NM  88003 \label{NMSU}}

\begin{abstract}

Because several observing programs are underway in various spectral
regimes to explore the Chandra Deep Field South (CDF--S), the value of
local photometric standards is obvious.  As part of an NOAO Surveys
Program to establish $u'g'r'i'z'$  standard stars in the southern
hemisphere, we have observed the central region of the CDF--S to
create local standards for use by other investigators using these
filters.  As a courtesy, we present the CDF--S standards to the public
now, although the main program will not finish until mid-2005.

\end{abstract}


\section{Introduction}

The photometric calibration of the Sloan Digital Sky Survey (SDSS) is
based on a relatively new, five-filter, wide-band photometric system
($u'g'r'i'z'$) defined by \citet{Fukugita96}.  This system offers
three astrophysical advantages over the established Johnson-Cousins
$UBVRI$ system: (1) sharper cutoffs of the band edges, (2) minimal
overlap of spectral regions between filters, and (3) filter breaks
chosen to exclude the strongest night sky emission lines.  We note the
SDSS $z'$ filter is open on the red end.  Therefore, the system
transformation coefficients are strongly dependent upon the choice of
detector that observers use to match the standard star network.

To support the calibration of the SDSS, \citet{Smith02} developed the
$u'g'r'i'z'$ standard star system using the U.S. Naval Observatory--
Flagstaff Station 1-m telescope, which is essentially a northern
hemisphere and equatorial network.  As part of the NOAO Surveys
Program,\footnote{{\tt http://www.noao.edu/gateway/surveys/programs.html}}
we have extended our work to establish $u'g'r'i'z'$  standard  stars
in the  southern hemisphere.  These new standards are tied to the
existing northern and equatorial network \citep{Smith02} developed for
the SDSS \citep{York00}.

The Great Observatories Origins Deep Survey (GOODS) \citep{DGG02} is a
multi-wavelength program which uses the great space-based, and some of
the  largest ground-based, observatories  to  obtain deep  imaging and
spectroscopy   of selected fields encompassing    both the Hubble Deep
Field North and the Chandra Deep Field South (CDF--S).  These fields
are among the most studied areas in the sky.  Several programs are
currently underway to probe the CDF--S  to an  unprecedented  depth in
several spectral regimes.  These include ground-based studies in the
optical \citep[e.g.][]{Arnouts01,Wolf01}, and the infrared
\citep[e.g.][]{Vandame02,Moy02}; and space-based studies in the X-ray
\citep[e.g.][]{Giacconi01,Giacconi02,Rosati02}.  Observations of this
field are also planned as part of the SIRTF Legacy and HST Treasury
Programs, the latter using the new Advanced Camera for Surveys (ACS).
This last program, the HST--ACS observations, is baselined to perform
$BVi'z'$ imaging \citep{Renzini02}; hence, $u'g'r'i'z'$ standard stars
would be immediately useful to the GOODS observers and others.

Until this report, no one has defined standard stars in the
$u'g'r'i'z'$ filter system in or near the CDF--S to use for
photometric calibration.  As part of our NOAO survey program, we have
observed the central region of the CDF--S to create local $u'g'r'i'z'$
standards, presented herein, to facilitate other investigators' use of
these filters.  Further, the placement of standard stars within the
CDF--S should ensure future observations, facilitating the study of
long-term time variable phenomena and transient events in the region.

In this paper, we present details of our observations in \S2, we
describe the data reductions in \S3, and we present the $u'g'r'i'z'$
magnitudes and colors of the CDF--S standard stars in \S4.  We will
examine remaining sources within the CDF--S in a future paper.

\section{Observations}

For this study we targeted the center of the CDF--S field at J2000
coordinates $\alpha =$ 03:32:28, $\delta = -$27:48:30
\citep{Giacconi01} [$l = 223.57; b = -54.44$].  Due to the nature of the
SDSS, most of the existing standard stars used to calibrate these
CDF--S local standards were on or near the equator.  We selected the
standard stars used as the basis for this work from the northern and
equatorial $u'g'r'i'z'$ network \citep{Smith02}, since they are
currently the only standard stars for this filter system.  We may use
additional standards being developed by our NOAO Survey Program for
future refinement of CDF--S field object magnitudes.

The data were collected with the CTIO 0.9-m telescope using the
Tek2k\#3 CCD operating at the cassegrain focus.  This ``Grade-1'' CCD
is thinned and has an anti-reflection coating, resulting in high
quantum efficiency similar to that of the detector used to establish
the initial standard system\footnote{{\tt
http://www.ctio.noao.edu/ccd\_info/ccd\_info.html}}.  Observers have
used this CCD in a stable configuration on this telescope since
October 1995.  The imager is controlled by Arcon software (version
3.3) and operated in multiple amplifier read mode.  The average gain
and read noise values for each of the amplifiers are listed in
Table~\ref{gains}.  The CCD has 24$\mu$ pixels which gives a scale of
0.396~arcsec/pixel and results in a 13.5 arc-minute field of view. We
observed with the CTIO SDSS $u'g'r'i'z'$ filter set.

\placetable{gains}

A current, machine-readable version of transmission curves for the
CTIO $u'g'r'i'z'$ filter set is not available.  For future use,
however, we have requested a full spectral transmission scan for this
filter set.  Likewise, a current machine-readable version of the CCD
spectral response is not available.  In the meantime, we have
generated preliminary response functions for the
CTIO-0.9m+Tek2k\#3+$u'g'r'i'z'$ filter system based upon (1) the
$u'g'r'i'z'$ filter transmission curves from the manufacturer (Custom
Scientific) for an identical filter set, (2) the CTIO Tek2k quantum
efficiency from the GIF plot at the CTIO CCD Information 
website\footnote{{\tt http://www.ctio.noao.edu/ccd\_info/ccd\_info.html}},
and the aluminum reflectances from \citet{Bennett63} as reproduced by
\citet{Kneale94}\footnote{{\tt http://www.gemini.edu/documentation/webdocs/spe/spe-te-g0043.pdf}} 
(we assume two aluminum reflecting surfaces in the system).
Machine-readable tables of these preliminary filter responses are
available at our public access URL\footnote{{\tt
http://www-sdss.fnal.gov:8000/$\sim$dtucker/Southern\_ugriz/index.html}},
where updated versions will be posted as new data become available.

\begin{figure}
\figurenum{1}
\includegraphics[angle=-90,scale=0.8]{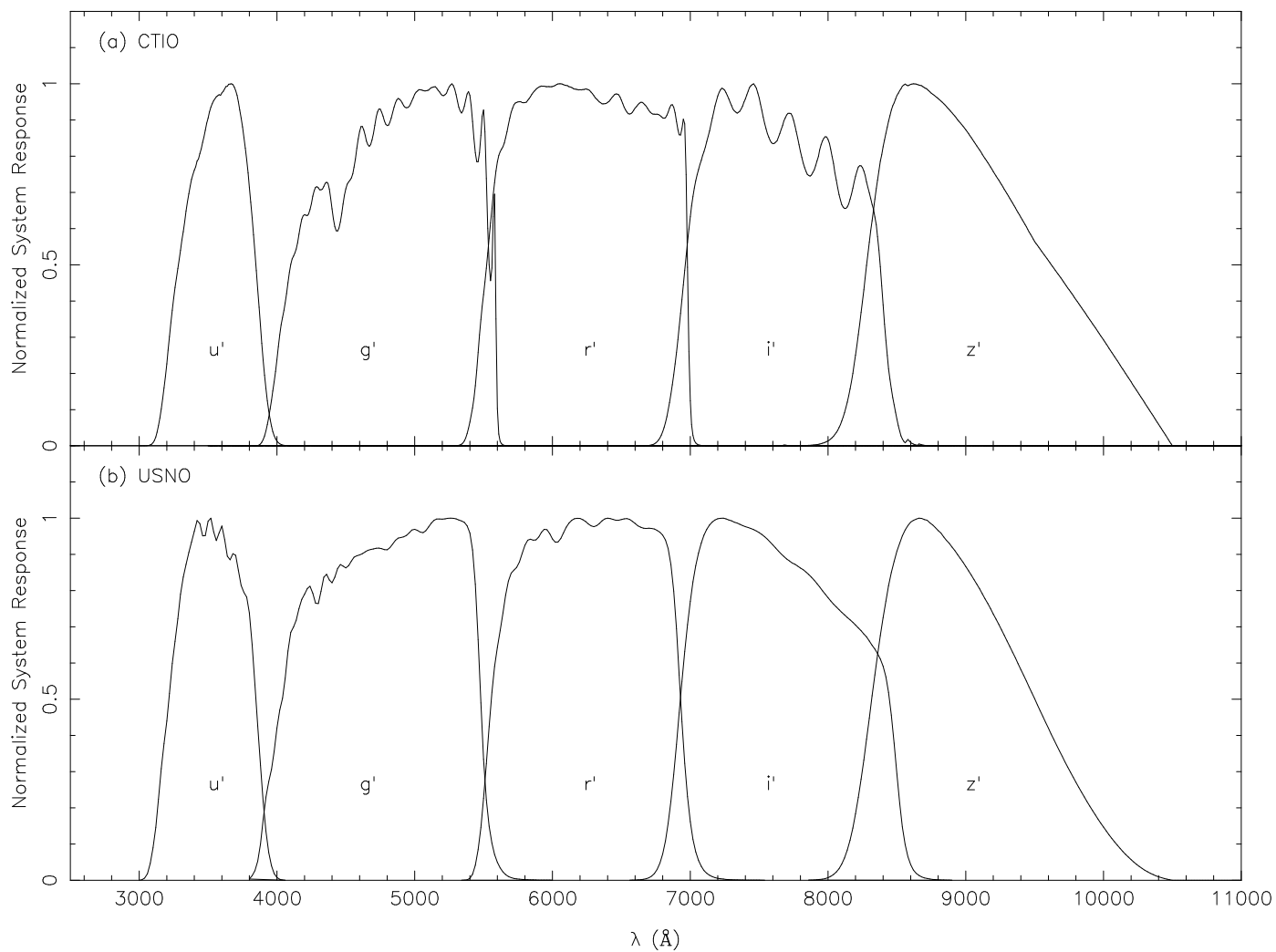}
\caption{
The (normalized) responses of the extra-atmospheric $u'g'r'i'z'$
system bandpasses for the (a) CTIO 0.9-m and the (b) USNO 1.0-m
telescope systems.  The CTIO data should be considered preliminary as
the CCD response function was taken from the GIF plot at {\tt
http://www.ctio.noao.edu/ccd\_info/ccd\_info.html} and is
approximately seven years old.  The filter data are from the manufacturer
for an identical filter set to the one used.  Note the similarity
between the two systems, supported by the low color term values
derived in the paper.  }
\label{response}
\end{figure}

In Figure~\ref{response} we plot these CTIO-0.9m+Tek2k\#3+$u'g'r'i'z'$
filter system responses and, for comparison, those from the
USNO-1.0m+Tek1k+$u'g'r'i'z'$ filter system used to set up the original
$u'g'r'i'z'$ standard star network.  The two system responses look
quite similar.  Given the uncertainties in calculating the CTIO
$u'g'r'i'z'$ response function, these curves are not inconsistent with
the results we report later in this paper (see \S~3 below).  The
values for the instrumental color terms we measure for our CTIO-0.9m
data are typically quite small -- ranging on average from about 0.02
to 0.06~mag per magnitude in color.  (We must emphasize, though, that
for the most accurate photometry --- i.e., systematic errors less than
a few percent --- instrumental color terms must be solved for and
applied when converting CTIO-0.9m $u'g'r'i'z'$ photometry to the USNO
standard system.)

\placefigure{response}

\begin{figure}
\figurenum{2}
\includegraphics[angle=0,scale=0.9]{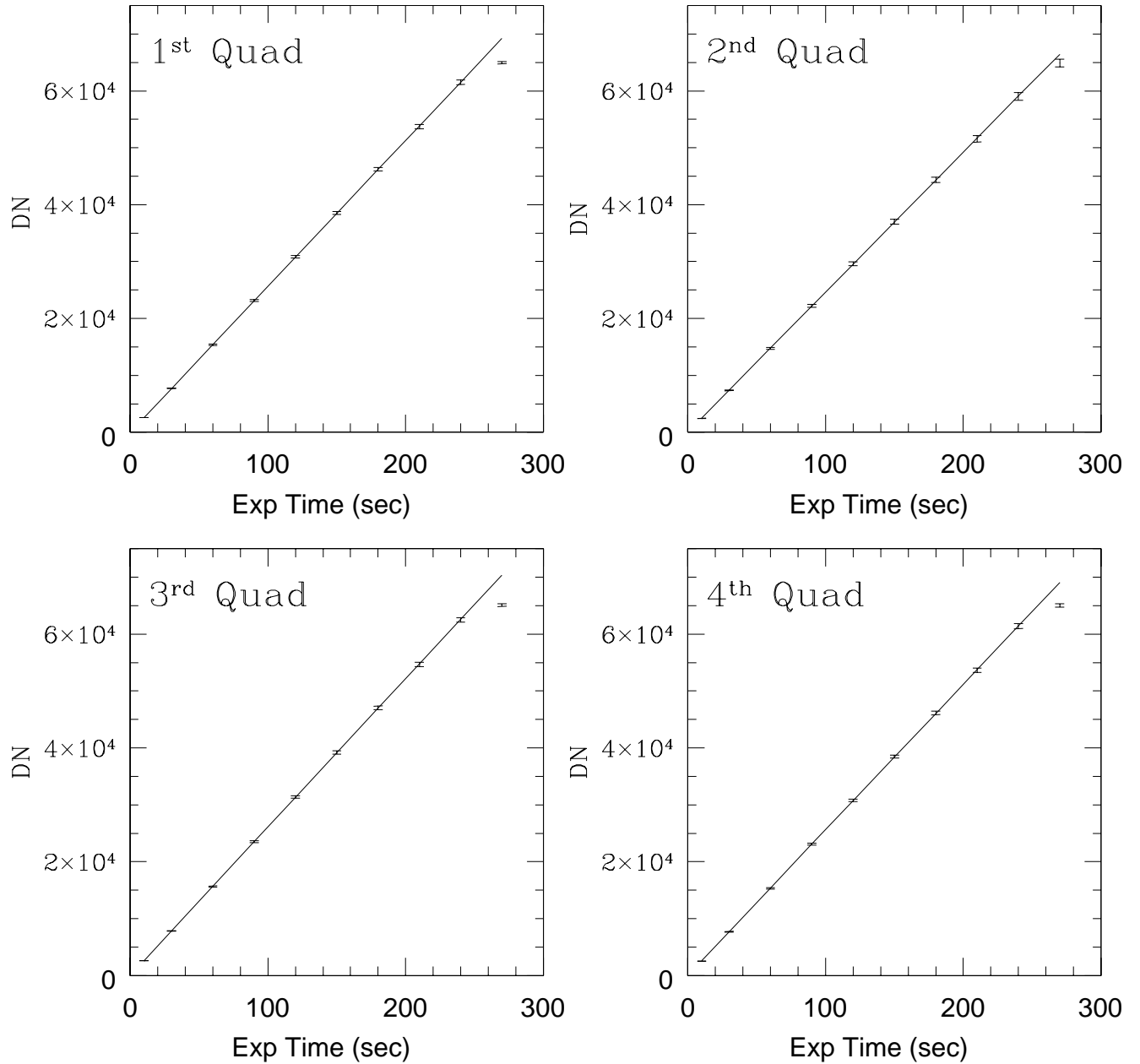}
\caption{
Linearity response tests for each quadrant of the Tek2k\#3 detector.
These show the weighted averages of three independent tests taken
during the May 2002 observing run.  As labelled, the quadrants refer
to: 1=LL; 2=UL; 3=LR; and 4=UR as viewed in the default orientation at
the telescope (pixel 0,0 in LL; 2048,2048 in UR).
}
\label{linearity}
\end{figure}

\begin{figure}
\figurenum{3}
\includegraphics[angle=0,scale=0.9]{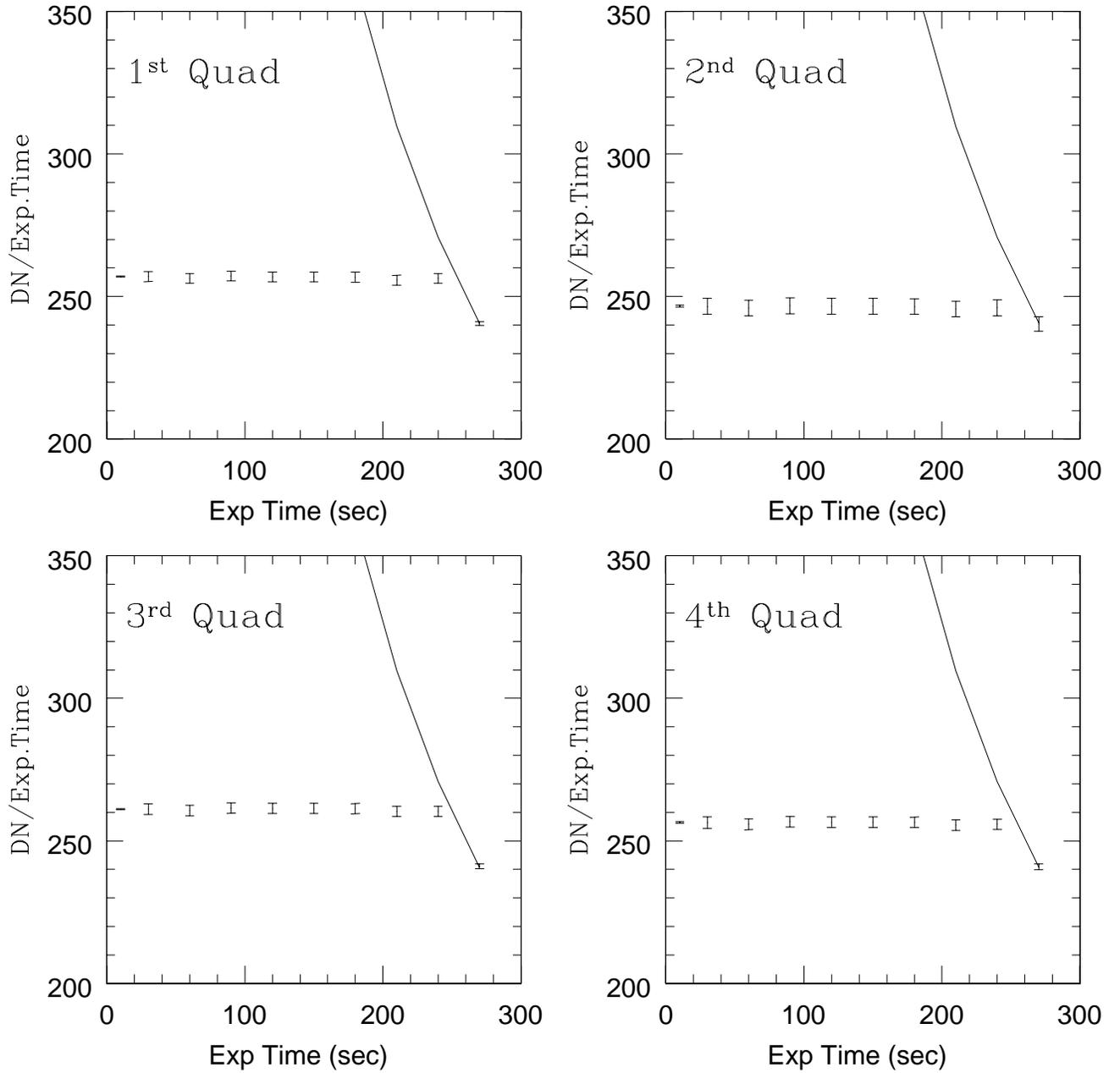}
\caption{
Deviation from linearity as a function of exposure time for the CCD.
These show the weighted averages for the same three independent tests
taken during the May 2002 observing run.  The solid line is the
calculated 65,000 DN line.  Labelling of the quadrants is the same
as in Figure 2.
}
\label{deviation}
\end{figure}

We have examined linearity of the CTIO system using the dome flat
lamps on different observing runs and found the response to be stable,
linear, and repeatable from 0$-\>$62,000 DN.  These tests are usually
performed once per observing run, weather ``permitting.''
Figure~\ref{linearity} shows the weighted average of the CCD response
as a function of exposure time for three separate linearity sequences
taken in May 2002.  Figure~\ref{deviation} gives the deviation from
linearity by exposure time for the same data.  Other tests show
identical responses.  The full results will be published in the final
paper at the end of the program.

\placefigure{linearity}

\placefigure{deviation}

We have also examined linearity (and shutter response) by moving a cluster
around the detector in a ``grid'' to look for measurement repeatability.
Preliminary results show no significant deviation of derived magnitudes
from these tests.  The supernovae monitoring group at CTIO has made its
shutter timing maps available to us, indicating expected deviations of
$\leq 0.12\%$ (1.2 milli-mags) from center to edge of the CCD for our
minimum exposure of five seconds.  Based on the shutter data obtained
by this group, the shutter exposure timing is stable and repeatable.

We collect and median-combine calibration frames daily, usually during
the afternoons.  These consisted of a minimum of 10 bias and dome flat
frames (10 per $g'r'i'z'$ filter).  The dome flats were obtained with
a color balance filter.  Because of a lack of photons, we did not
obtain $u'$ dome flats.  The dome flat images help us monitor the
status of the CCD and look for changes in the flat-field structure.
In addition, twilight sky flats were collected in all five filters
during one or both of the twilight periods on each observable night.
These were median-combined at the end of each observing run to produce
a ``master'' twilight flat and used in the reduction of the data
frames.  We chose this approach to maintain consistency with the
original standard network.  At some point during each observing run,
we usually collected long dark frames to monitor changes in the hot
pixels on the CCD and to look for light leaks.  We generated fringe
correction frames using the long program object exposures.  These were
applied to the $i'$ and $z'$ band images.

During a typical night in our standards program, we observe five or
six existing standard fields three times each --- at the start, near
the middle, and at the end of the night --- in order to establish an
extinction and color term baseline.  Between these extended standard
sequences, we usually alternate one or two program fields and one to
three standard fields.  We use these to monitor the extinction values
established by the longer extinction scans.  This observing method
allows us to maximize the number of target fields while continuously
monitoring the atmosphere for changes.  Exposure times for the
established standard fields ranged from 5 to 240 seconds, with a mean
of 7.9 seconds for the shortest exposures (generally the $g'r'i'$
filters).

The program observations --- performed under apparently clear conditions
on nine different nights spanning three separate observing runs in
2001~September, 2002~February, and 2002~October --- consisted of two
separate exposure cycles, resulting in eighteen data points per filter.
We obtained two additional $r'$ band observations under obviously
non-photometric conditions during the September 2001 observing run.
These later observations were obtained to use in a differential search
for short period variable stars but were not included in the calculations
of the final magnitudes.  After processing, we determined that our
observations from 2002~October~5 were not photometric, so we discarded
them from our final calculations of the calibrated magnitudes, leaving
a total of sixteen photometric data points per filter.
Table~\ref{obs} lists the circumstances of all our observations of the
CDF--S field.  The first two columns give  the UT  and Modified Julian
Date (MJD\footnote{The Modified Julian Date is defined by the relation
MJD $\equiv$ JD $-$  2400000.5, where JD is the Julian Date.})  of the
observation; the third column gives the approximate airmass at the
start of  the observation sequence.  The exposure times (in seconds)
for each filter appear in column four, and the last column gives the
observer impression and reduction decisions concerning the sky
conditions during the observations.

\placetable{obs}

\section{Reductions}

We performed reductions using version {\tt v8.0} of the SDSS software
pipeline {\tt mtpipe} \citep[see][]{Tucker03}, an earlier version of
which was used in the setup of the original $u'g'r'i'z'$ standard star
network \citet{Smith02}.  This pipeline consists of four main packages:

\begin{itemize}

\item {\tt preMtFrames}, which creates the directory structure for the
	reduction of a night's data, including parameter files needed
	as input for the other three packages, and runs
	quality-assurance tests on the raw data.

\item {\tt mtFrames}, which processes the images and performs object
	detection and aperture photometry on target field images.
	The processing steps include zero subtraction, flat-field
	and fringe-frame correction.

\item {\tt excal}, which takes the aperture photometry lists for the
	standard star target fields (i.e., stars from
	\citealt{Smith02}), identifies the individual standard stars
	within those fields, and fits the observed raw counts and
	known  $u'g'r'i'z'$ magnitudes to a set of photometric
	equations to obtain extinction and zero point coefficients.
	The output from this package allows us to monitor the
	stability of the night.  The default analysis block is
	three hours, but can be changed as required based upon the
	data present and upon trends in the reductions.  A minimum of
	ten standard stars are required for the night to be useable.

\item {\tt kali}, which applies the fitted photometric equations to
	the aperture photometry lists of program target fields for
	the appropriate analysis block (e.g., the CDF--S field).

\end{itemize}

We note  a few small  differences between the  methods employed in the
current   reductions  and  those  used  in   setting  up  the original
$u'g'r'i'z'$ standard  star network of  \citet{Smith02}.  First, since
we tailored our  current effort towards calibrating standard  stars
which are typically much  fainter  than the \citet{Smith02}  standards
(which were generally in the  range $r' \approx 8$  -- 12), we chose a
smaller extraction size for our aperture photometry to reduce or
minimize the background sky contribution to the noise.  For the
\citet{Smith02}    standards,   we  employed   a  24\arcsec-diameter
aperture in order to avoid problems associated with defocusing the
brightest stars (required for some the observations).  In the current
program, we have chosen a 14.86\arcsec-diameter aperture.  This
smaller aperture reduces the effects of sky noise for the fainter
CDF--S target stars; as an added bonus, this size is the one used in
the photometric calibration of the SDSS 2.5m data
\citep{GCRS98,LGS99,York00,Sto02}.  Tests on the fainter standards in
\citet{Smith02} show no significant deviations from the published
magnitudes using this smaller extraction aperture.

Second, the current version of {\tt mtpipe} uses photometric equations
which are slightly modified from the form described in \S4.2 of
\citet{Smith02}.   The photometric  equations  employed in the current
paper are the following:

\begin{eqnarray}
u'_{\rm inst} & = & u'_{\rm o} + a_{u} + k_{u} X \nonumber \\
              &   & + b_{u} [(u'-g')_{\rm o} - (u'-g')_{\rm o,zp}] \nonumber \\
              &   & + c_{u} [(u'-g')_{\rm o} - (u'-g')_{\rm o,zp}] [X-X_{\rm zp}]\;, \\
g'_{\rm inst} & = & g'_{\rm o} + a_{g} + k_{g} X \nonumber \\
              &   & + b_{g} [(g'-r')_{\rm o} - (g'-r')_{\rm o,zp}] \nonumber \\
              &   & + c_{g} [(g'-r')_{\rm o} - (g'-r')_{\rm o,zp}] [X-X_{\rm zp}]\;, \\
r'_{\rm inst} & = & r'_{\rm o} + a_{r} + k_{r} X \nonumber \\
              &   & + b_{r} [(r'-i')_{\rm o} - (r'-i')_{\rm o,zp}] \nonumber \\
              &   & + c_{r} [(r'-i')_{\rm o} - (r'-i')_{\rm o,zp}] [X-X_{\rm zp}]\;, \\
i'_{\rm inst} & = & i'_{\rm o} + a_{i} + k_{i} X \nonumber \\
              &   & + b_{i} [(i'-z')_{\rm o} - (i'-z')_{\rm o,zp}] \nonumber \\
              &   & + c_{i} [(i'-z')_{\rm o} - (i'-z')_{\rm o,zp}] [X-X_{\rm zp}]\;, \\
z'_{\rm inst} & = & z'_{\rm o} + a_{z} + k_{z} X \nonumber \\
              &   & + b_{z} [(i'-z')_{\rm o} - (i'-z')_{\rm o,zp}] \nonumber \\
              &   & + c_{z} [(i'-z')_{\rm o} - (i'-z')_{\rm o,zp}] [X-X_{\rm zp}] .
\end{eqnarray}

Taking  the $g'$ equation as an  example, we note that $g'_{\rm inst}$
is the measured  instrumental    magnitude,   $g'_{\rm o}$   is    the
extra-atmospheric     magnitude,    $(g'-r')_{\rm    o}$      is   the
extra-atmospheric color, $a_{g}$ is the nightly zero point, $k_{g}$ is
the first order  extinction    coefficient,  $b_{g}$ is the     system
transform coefficient, $c_{g}$ is the  second order (color) extinction
coefficient, and $X$ is the airmass of the observation.  The zeropoint
constants,   $X_{\rm zp}$   and  $(g'-r')_{\rm   o,zp}$  were defined,
respectively,  to be  the  average  standard star observation  airmass
$<X>$  =   1.3 and the   ``cosmic color,''  as   listed in  Table~3 of
\citet{Smith02}.    Note that the   above  equations differ from their
analogs in \citet{Smith02} by the inclusion of zeropoint colors in the
system transform  (``b'')  terms.   (Note  also  that  there  are  some
differences in the calibration methodology used in the current paper as
opposed to that now used in standard photometric  calibrations  of the
SDSS imaging data.  In particular, standard  SDSS calibrations now use
different values for the    zeropoint colors; further,  standard  SDSS
calibrations  now  index  the $i'$  filter   to $(r'-i')$ and   not to
$(i'-z')$; for more details see \citet{Tucker03}.)

Third, in \citet{Smith02}, since we used one telescope (the USNO 1-m)
for all the observations in setting up  the original  $u'g'r'i'z'$
standard star network, we set all values of the system transform (``b'')
coefficients identically to zero.  Here, we are using a different
telescope, so we solve for these ``b'' terms.

Fourth, instead of using the first-order inverse photometric equations
to convert from instrumental   magnitudes to calibrated magnitudes  in
{\tt kali} (eqs. 9 -- 13 of \citealt{Smith02}), the current version of
{\tt mtpipe}  does this conversion  by solving the above equations
iteratively.

Finally, since none of the $(u'-i')$ and $(u'-z')$ colors of the final
set of CDF--S  standards are  very red,  no red leak  corrections were
applied to the CDF--S $u'$ magnitudes.

With these caveats in mind, the night characterization data from {\tt
mtpipe} for each of the photometric nights included in this project
are given in Table~\ref{coeffs}.  These data include the MJD of
observation (column 1), filter (column 2), zero points (column 3),
system transformation terms (column 4), and first-order extinction
terms (columns 5 through 7).  Note that the zero
points and system tranformation terms are solved on a night-by-night
basis; since it is not uncommon for the first-order extinctions to vary
during a night, we typically solve for them in 3-hour-long blocks of time.
Finally, columns 8 and 9 give the rms errors for, and numbers of, the
standard stars observed that night which were used in the photometric
solutions.  The weighted mean averages are listed by filter at the
bottom of the table as an aid to observers looking for mean site
values.  In a footnote, we also list the second order extinction terms
derived in \citet{Smith02}.

\placetable{coeffs}

\begin{figure}
\figurenum{4}
\includegraphics[angle=0,scale=0.9]{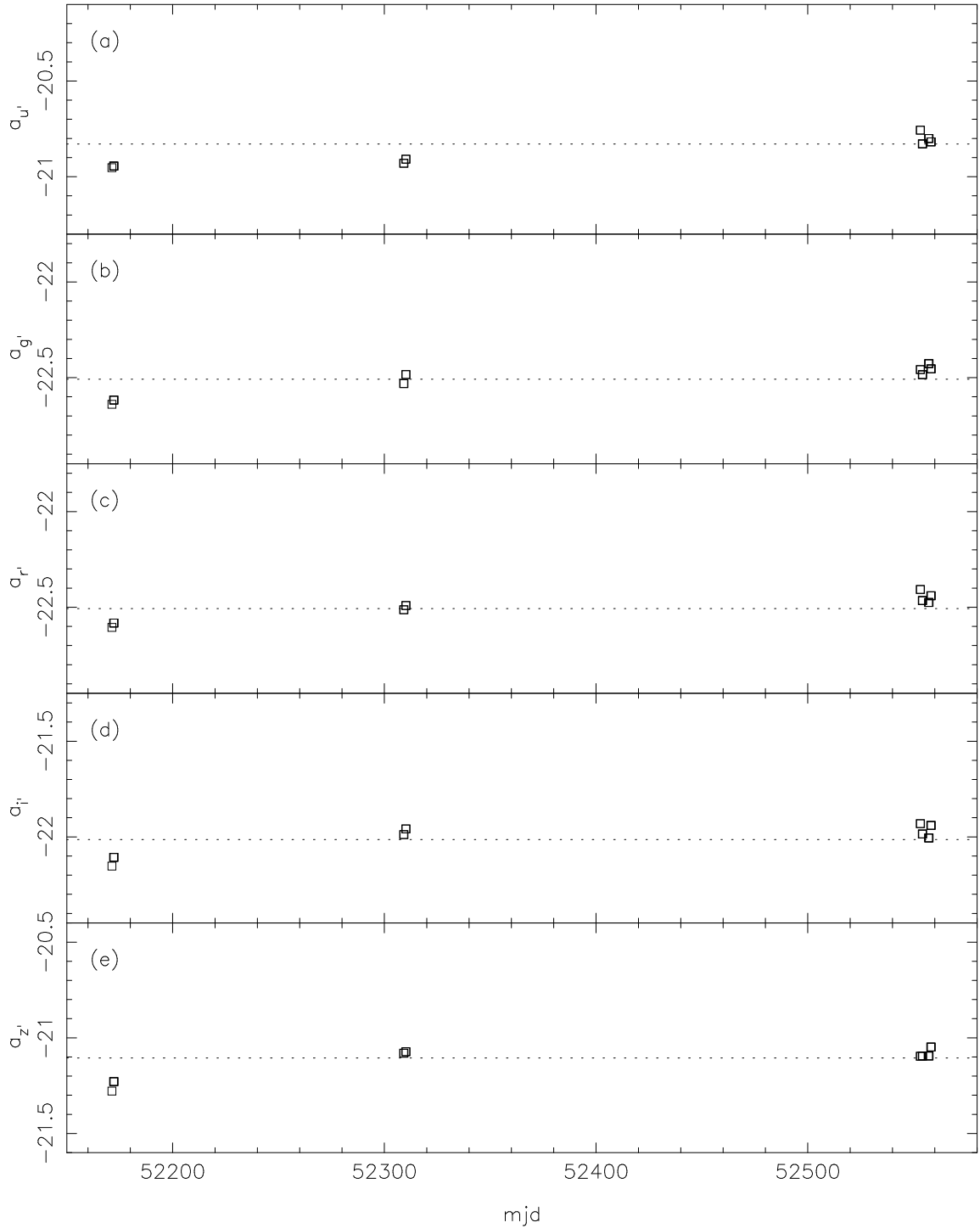}
\caption{
The photometric zeropoints for each night of data, by filter,
$u'g'r'i'z'$, from top to bottom. The dotted line indicates the mean
zeropoints.  Note the slight degradation of telescope throughput
with time.  These values were taken from Table~\ref{coeffs}.
}
\label{Zplot}
\end{figure}

\begin{figure}
\figurenum{5}
\includegraphics[angle=0,scale=0.9]{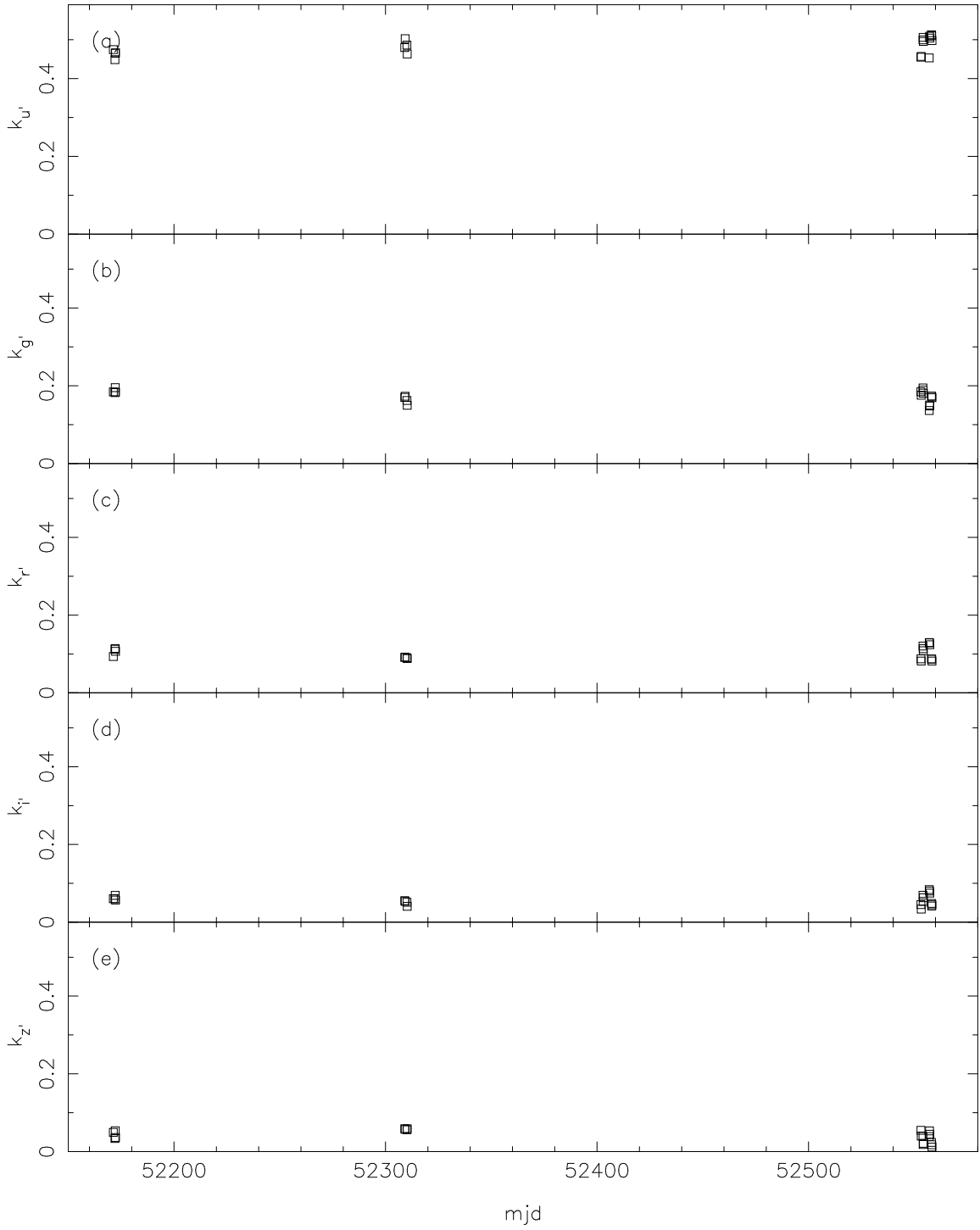}
\caption{
The first order extinction coefficients for each reduction block by
filter, $u'g'r'i'z'$, from top to bottom.  These values were taken
from Table~\ref{coeffs}.
}
\label{kplot}
\end{figure}

\begin{figure}
\figurenum{6}
\includegraphics[angle=0,scale=0.9]{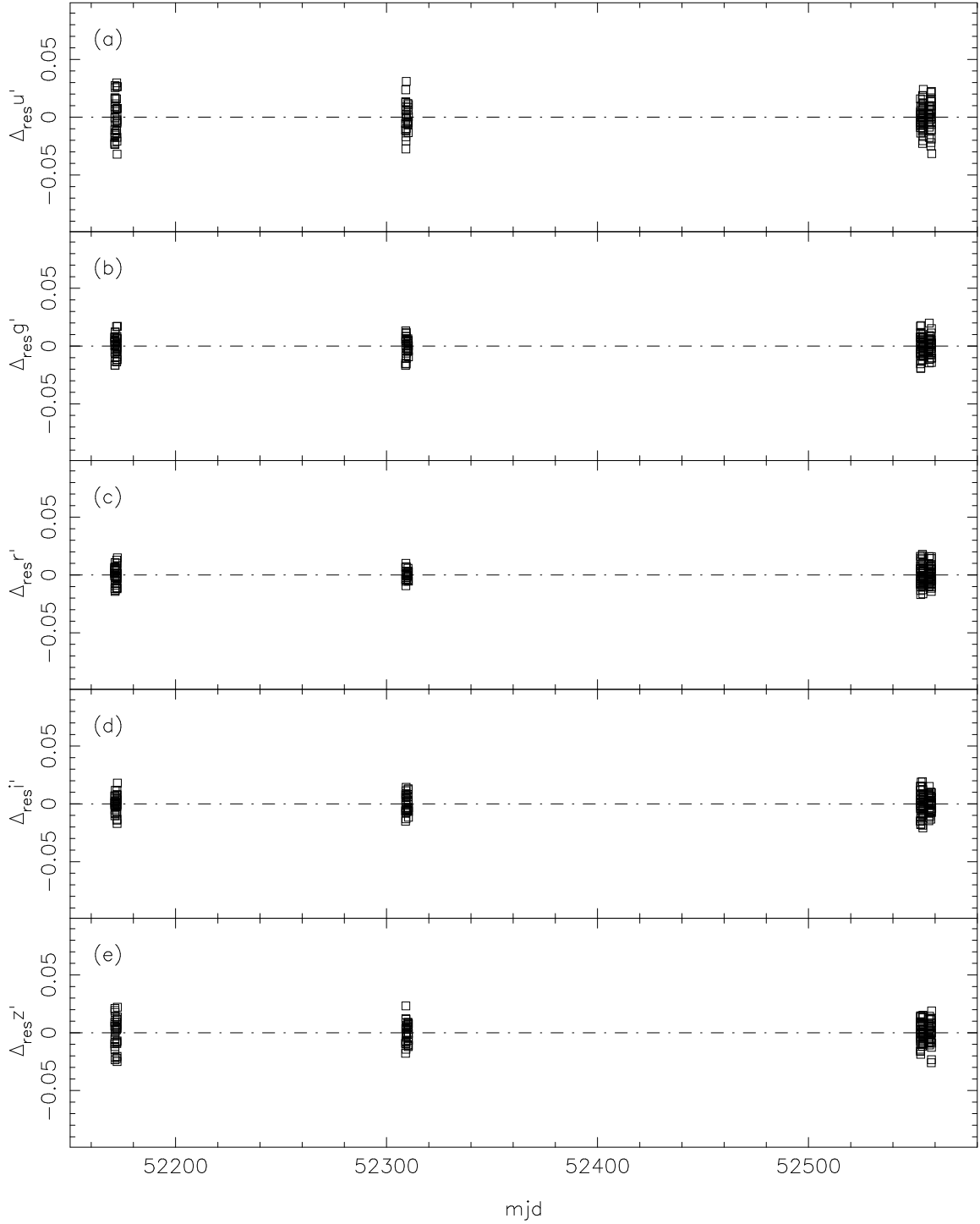}
\caption{
The residuals of the {\tt excal} solution for the standard stars used
in the reductions by filter, $u'g'r'i'z'$, from top to bottom.  These
are plotted in the sense of (observed$-$standard) for each night's
solution.
}
\label{std-rms}
\end{figure}

Figure~\ref{Zplot} shows the photometric zeropoint versus MJD for each
filter.  We see the slight degradation of telescope throughput with
time, a result of the mirror not being re-aluminized over the course
of this program to date.  Figure 5 of \citet{Smith02} shows similar
trends for the USNO-1.0m telescope, but the effect of re-aluminization
is clearly seen.  Figure~\ref{kplot} shows the first order extinction
coefficients for each reduction block by MJD.  As shown, all the
nights used for these data were well behaved.  Finally,
Figure~\ref{std-rms} shows the residuals of the {\tt excal} solution
for each filter by MJD for each of the standard stars used in the
photometric solution.  The plot is in the sense (observed $-$ true),
where ``true'' comes from
\citet{Smith02}.  This plot may be slightly misleading since we did
night-by-night solutions rather than a global solution.

\placefigure{Zplot}

\placefigure{kplot}

\placefigure{std-rms}

At this point in the reduction process using {\tt mtpipe}, we had
sixteen calibrated object lists for the CDF--S, one list for each of
the sixteen photometric exposures of this field.  We combined these
lists by taking the (unweighted) mean magnitude of each object in each
filter.  To avoid problems associated with signal-to-noise mismatches
between the long and the short exposures, we only included in the mean
magnitudes those measurements having photon noise errors of
$\leq 0.05$~mag.  We excluded saturated measurements from the mean
magnitude calculations.  The resulting list of candidate CDF--S
standards contains 355 objects.

We culled this list using the following criteria:
\begin{itemize}

\item The mean magnitude in $r'$ must have been derived from at least
	ten good individual measurements.

\item The standard deviation of the individual measurements in $r'$
	must be less than 0.10~mag (to avoid variables).

\item The error in the mean magnitude in $r'$ (standard deviation of
	the mean) must be less  than  0.03~mag (to be  useful  as a
	standard star).

\item The mean magnitude in $r'$ must be less than 18.0, which is
    	approximately the limit of our present  data  to achieve an
    	error in the mean $r'$ magnitude of less than 0.03~mag.

\end{itemize}

After culling the {\tt mtpipe} output using the above criteria, only
24 objects remained as candidate standard stars.  Then, we ran SExtractor
\citep{SExtractor} on one of our long CDF--S exposures to obtain the
automated star-galaxy classifier value.  Any object classified as
non-stellar either by SExtractor, or by eye, was removed from the list
of candidate standards.  This resulted in removal of two galaxies.
Further, we used the ESO Imaging Survey (EIS) stellar catalog
\citep{Groen02} to confirm our star-galaxy separation for the brighter
objects in our frames, and we deferred to the much deeper EIS catalog
classification for our fainter sources.  Finally, we performed a
coordinate match with the COMBO-17 Survey \citep{Wolf01} $BVR$ sources
to obtain cross reference designations.  The final list presented here
contains 22 CDF--S standard stars.

\section{The CDF--S Standard Stars}

Here, we present the calibrated magnitude and color data for each star
in our final list of CDF--S local standards.

\begin{figure}
\figurenum{7}
\includegraphics[angle=-90,scale=1.2]{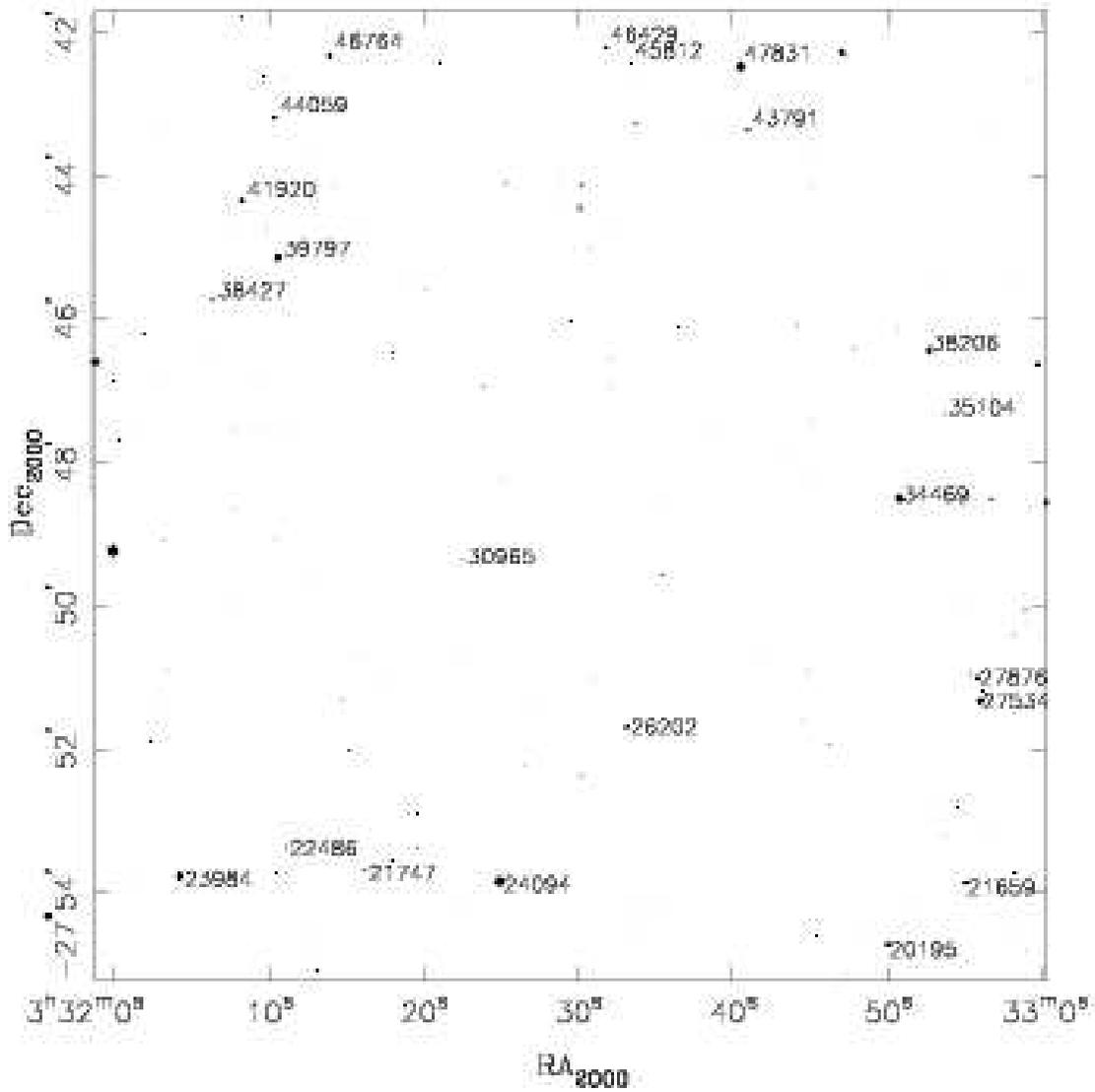}
\caption{
Finder chart for the CDF--S field, an $r'$ band image.  The marked stars
are those selected to act as local standards.  The numbers are the
COMBO-17 star designations from Table~4.  Coordinates for the center of
the field are $\alpha$ = 03:32:28, $\delta = -$27:48:30, J2000.
}
\label{cdfs-finder}
\end{figure}

Table~\ref{stdstars} shows the CDF--S standard stars arranged in order
of increasing $r'$-band magnitude and contains the COMBO-17 designation
and the right ascension and declination (J2000) in the first three
columns.  The next five columns give the $r'$ band magnitudes and four
color indices.   These five columns are linked with the following five
columns (9-13) which give the estimated rms error --- i.e., the
standard deviation of the mean ---  of the measurements.   As a note,
during  the reductions we  calculated the five  filter magnitudes.  We
report  colors  here   as  an   observational   aid.   The  associated
uncertainties for  the  colors are derived   from the magnitude errors
added   in quadrature.  As  such,  they may be slightly overestimated,
since  magnitude errors  in  different filters  tend to be correlated.
The last five columns  of  this table  list the number  of  individual
measurements, by  filter,  that  were used   to  determine  the  final
magnitudes.  Finally, Table~\ref{crossref} gives the COMBO-17  and ESO
Imaging Survey (EIS) \citep{Arnouts01} designations with the coordinates
(J2000) for each of the stars in the final standard list.  As with
Table~\ref{stdstars}, this list is arranged in order of increasing
$r'$-band magnitude.  A finder chart, based upon on one of our long
$r'$ band images and showing the location of each of these standard
stars, appears in Figure~\ref{cdfs-finder}.

\placetable{stdstars}

\placetable{crossref}

\placefigure{cdfs-finder}

\begin{figure}
\figurenum{8}
\includegraphics[angle=0,scale=1.4]{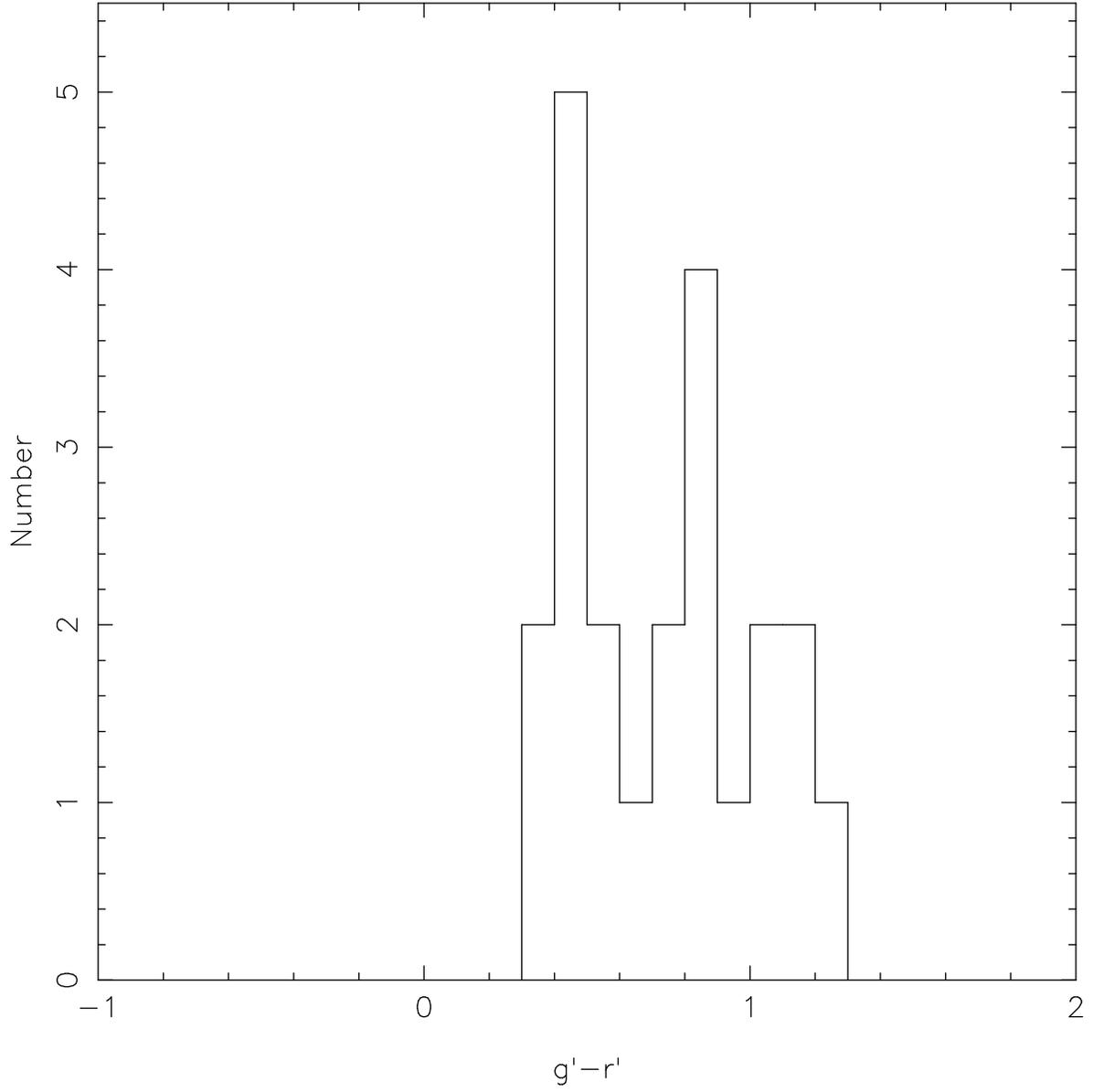}
\caption{
The $(g'-r')$ color distribution of local standards in the CDF--S field.
}
\label{histogram}
\end{figure}

\begin{figure}
\figurenum{9}
\includegraphics[angle=0,scale=0.9]{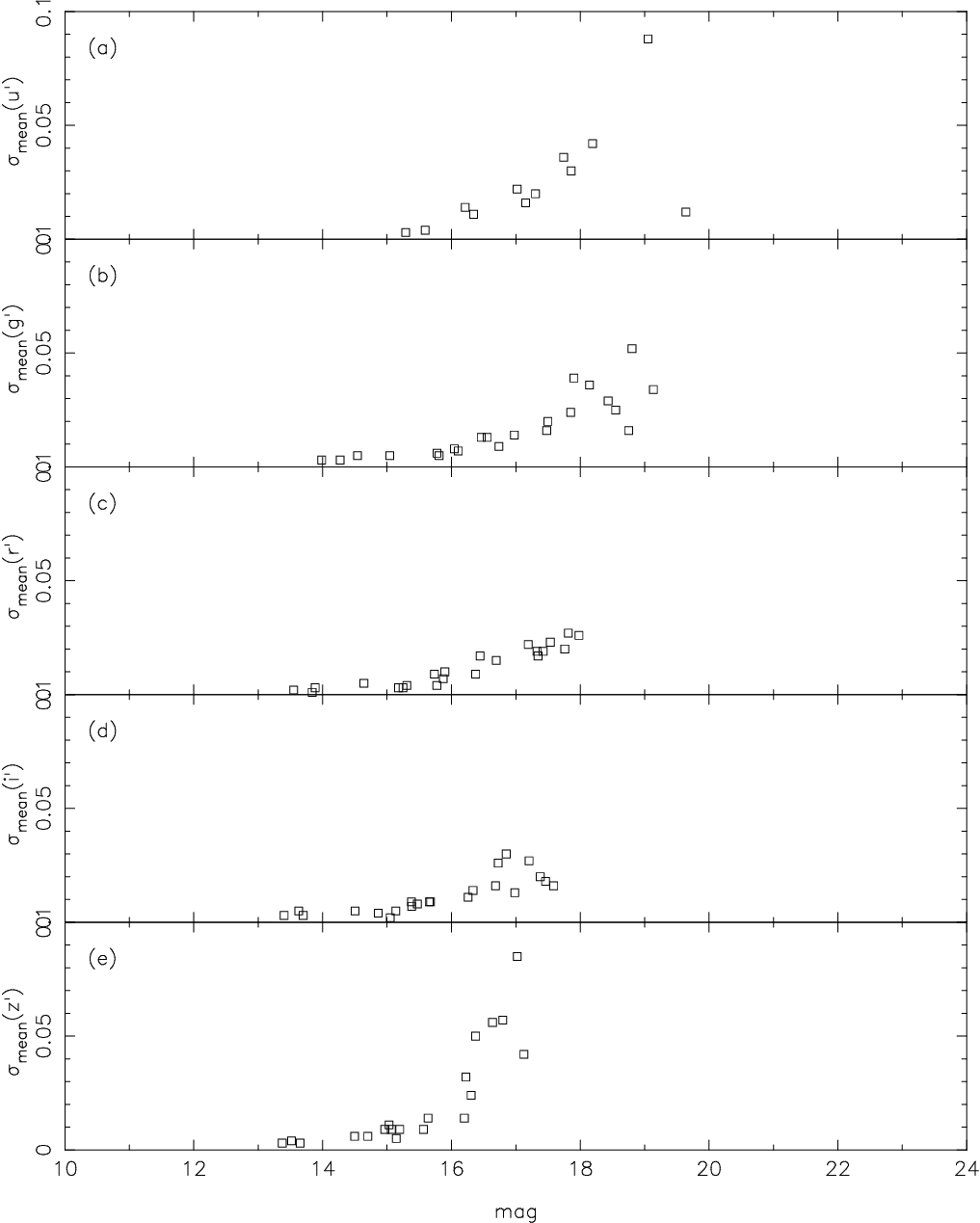}
\caption{
The estimated rms error in the calibrated magnitudes
for the local standards in the CDF--S field versus magnitude:
(a) $\sigma_{\rm mean}(u')$ vs.\ $u'$,
(b) $\sigma_{\rm mean}(g')$ vs.\ $g'$,
(c) $\sigma_{\rm mean}(r')$ vs.\ $r'$,
(d) $\sigma_{\rm mean}(i')$ vs.\ $i'$, and
(e) $\sigma_{\rm mean}(z')$ vs.\ $z'$.
These values were taken from Table~\ref{stdstars}.
}
\label{mag-sig}
\end{figure}

\begin{figure}
\figurenum{10}
\includegraphics[angle=0,scale=0.9]{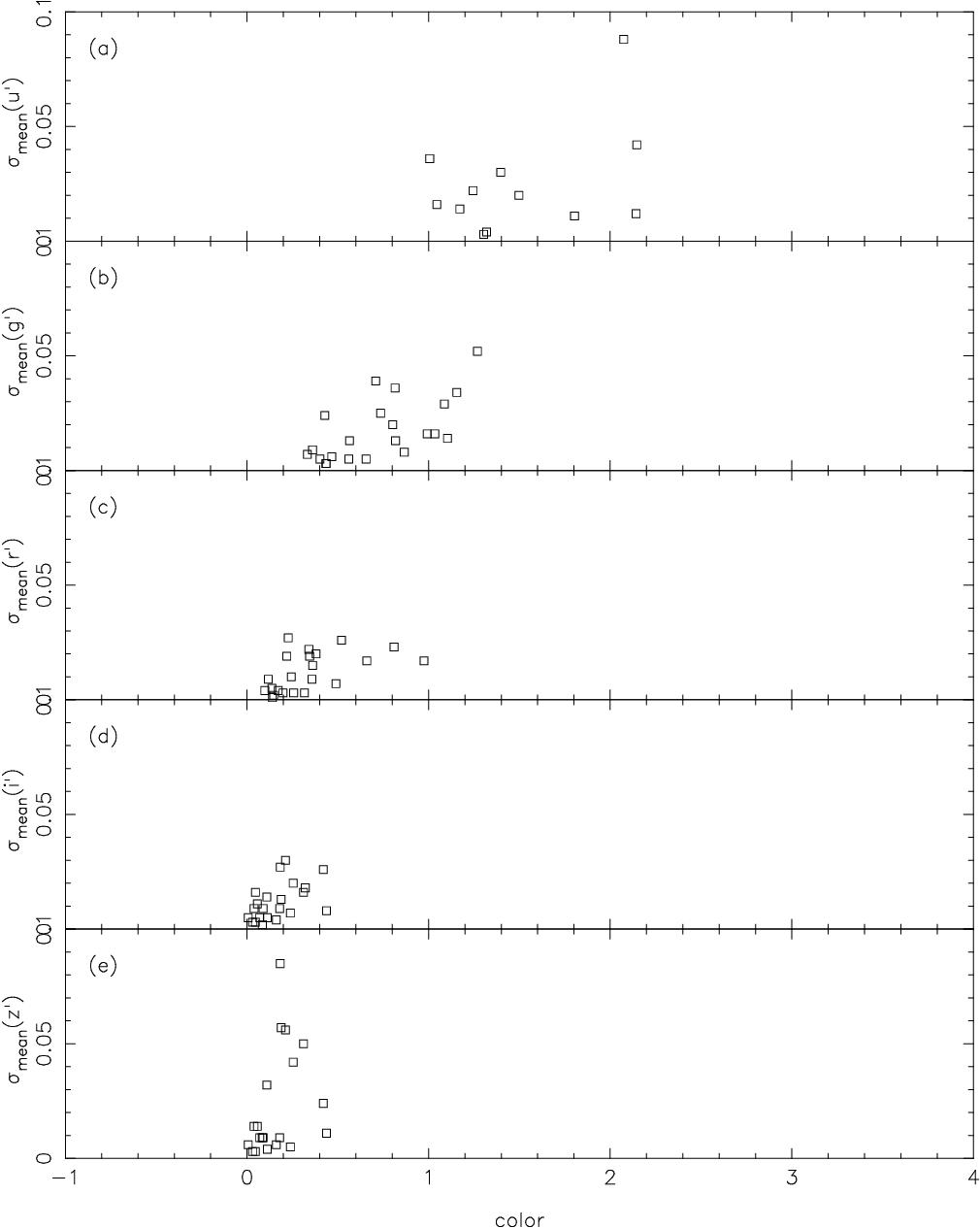}
\caption{
The estimated rms error in the calibrated magnitudes
for the local standards in the CDF--S field versus magnitude:
(a) $\sigma_{\rm mean}(u')$ vs.\ $(u'-g')$,
(b) $\sigma_{\rm mean}(g')$ vs.\ $(g'-r')$,
(c) $\sigma_{\rm mean}(r')$ vs.\ $(r'-i')$,
(d) $\sigma_{\rm mean}(i')$ vs.\ $(i'-z')$, and
(e) $\sigma_{\rm mean}(z')$ vs.\ $(i'-z')$.
These values were taken from Table~\ref{stdstars}.
}
\label{color-sig}
\end{figure}

We present a histogram of the  distribution of $(g'-r')$ colors of the
CDF--S local standards in Figure~\ref{histogram}.  Figure~\ref{mag-sig}
shows the estimated rms errors (the standard deviation of the  mean) in
the calibrated magnitudes versus the calibrated magnitude for the CDF--S
local standards in each of the five filters.  Figure~\ref{color-sig}
shows the rms errors for the calibrated magnitudes versus the color for
the CDF--S local standards in each of the filters.

\placefigure{histogram}

\placefigure{mag-sig}

\placefigure{color-sig}

We began this effort in September 2000 using the 0.9-m at CTIO and the
same  observers and reduction software that  were used in the setup of
the  original     $u'g'r'i'z'$   standard  network    described     in
\citet{Smith02}.  We undertook this effort to provide a uniform set of
standard stars in the $u'g'r'i'z'$ system across the sky with the goal
to provide future investigators a convenient starting grid to establish
tertiary  standards   for their  own work,  without   the  need for an
extensive end-to-end standardization effort.  Observations of this
field will continue through the course of our survey  program.
As they become available, updated magnitudes and colors, along with all
of the southern standard stars, will be posted our a public access URL
mentioned previously.
We estimate that the entire grid of standard star from this project will
become available for public dissemination by mid-2005.

\acknowledgments

We acknowledge the  National Optical  Astronomy Observatories for  the
observing time  granted through the  NOAO Survey Program, the staff at
CTIO for its help, especially Edgardo Cosgrove, Arturo Gomaez, Nick
Suntzeff, and Stefanie Wachter.  We are grateful to the National Science
Foundation for support through AST-0098401.  JAS acknowledges Amy Felty
and Jeanne Odermann for their careful reading of the manuscript and
proper application of the laws of English.

DLT was supported  by the US  Department of Energy under contract No.\
DE-AC02-76CH03000.

We thank the anonymous referee for insightful comments which
stregthened this paper and enhanced the overall presentation of the
data.  In this vein, we also extend our thanks to all manuscript
reviewers for their diligent work.

The EIS data  are based on observations  carried out using the MPG/ESO
2.2m Telescope and the  ESO New Technology  Telescope (NTT) at  the La
Silla observatory under Program-ID No. 164.0-0561.

This research has made use of the SIMBAD database, operated at CDS; the
VizieR catalog access tool, CDS; and the Aladin, developed by CDS,
Strasbourg, France.


\clearpage

\clearpage

\begin{center}
\begin{deluxetable}{lcc}
\footnotesize
\tablecaption{CCD Parameters\label{gains}}
\tablewidth{0pt}
\tablehead{
\colhead{Side} & \colhead{Gain (epadu)}   & \colhead{Read Noise (e$^{-}$)}
}
\startdata
upper Left  &  3.1  &  4.7  \\
lower Left  &  3.0  &  5.4  \\
upper Right &  3.0  &  4.6  \\
lower Right &  3.0  &  5.1
\enddata
\end{deluxetable}
\end{center}

\clearpage

\begin{deluxetable}{ccccccccl}
\tablecaption{Observing Circumstances for the CDF--S\label{obs}}
\tablehead{
  \colhead{YYMMDD} &
  \colhead{MJD} &
  \colhead{Airmass} &
  \multicolumn{5}{c}{Exposure (sec)} &
  \colhead{Comments} \\
  \colhead{} &
  \colhead{} &
  \colhead{} &
  \colhead{ $r'$ } &
  \colhead{ $g'$ } &
  \colhead{ $u'$ } &
  \colhead{ $i'$ } &
  \colhead{ $z'$ } &
  \colhead{}
}
\startdata
010919 & 52171 & 1.04    & 30  & 30  & 240  & 30  & 45    & Photometric \\
       &       & 1.03    & 180 & 180 & 1440 & 180 & 270   & Photometric \\
010920 & 52172 & 1.02    & 30  & 30  & 240  & 30  & 45    & Photometric \\
       &       & 1.02    & 180 & 180 & 1440 & 180 & 270   & Photometric \\
010922 & 52174 & \nodata & 30  & \nodata & \nodata &  \nodata &   \nodata   & Differential ($r'$ filter only) \\
       &       & \nodata & 180 & \nodata & \nodata &  \nodata &   \nodata   & Differential ($r'$ filter only) \\
       &       & \nodata & 30  & \nodata & \nodata &  \nodata &   \nodata   & Differential ($r'$ filter only) \\
       &       & \nodata & 180 & \nodata & \nodata &  \nodata &   \nodata   & Differential ($r'$ filter only) \\
020204 & 52309 & 1.11    & 30  & 30  & 240  & 30  & 45    & Photometric \\
       &       & 1.13    & 180 & 180 & 1440 & 180 & 270   & Photometric \\
020205 & 52310 & 1.17    & 30  & 30  & 240  & 30  & 45    & Photometric \\
       &       & 1.23    & 180 & 180 & 1440 & 180 & 270   & Photometric \\
021005 & 52552 & 1.01    & 20  & 25  & 150  & 20  & 30    & Non-photometric on analysis\\
       &       & 1.00    & 180 & 180 & 1440 & 180 & 240   & Non-photometric on analysis\\
021006 & 52553 & 1.02    & 20  & 25  & 150  & 20  & 30    & Photometric \\
       &       & 1.02    & 180 & 180 & 1440 & 180 & 240   & Photometric \\
021007 & 52554 & 1.02    & 20  & 25  & 150  & 20  & 30    & Photometric \\
       &       & 1.01    & 180 & 180 & 1440 & 180 & 240   & Photometric \\
021010 & 52557 & 1.01    & 20  & 25  & 150  & 20  & 30    & Photometric \\
       &       & 1.01    & 180 & 180 & 1440 & 180 & 240   & Photometric \\
021011 & 52558 & 1.05    & 20  & 25  & 150  & 20  & 30    & Photometric \\
       &       & 1.04    & 100 & 125 & 750  & 100 & 150   & Photometric \\
\enddata
\end{deluxetable}

\clearpage

\begin{deluxetable}{lcccccccc}
\tabletypesize{\tiny}
\tablecaption{Night Characterization Coefficients \& Averages\tablenotemark{*}
\label{coeffs}}
\tablewidth{0pt}
\tablehead{
  \colhead{MJD} &
  \colhead{Filter} &
  \colhead{Zeropoint (a)} &
  \colhead{Instr. Color (b)} &
  \multicolumn{3}{c}{1st-Order Ext. (k)} &
  \colhead{Std. rms} &
  \colhead{\# Std.} \\
  \colhead{}  &
  \colhead{}  &
  \colhead{}  &
  \colhead{}  &
  \colhead{block 0} &
  \colhead{block 1} &
  \colhead{block 2} &
  \colhead{(mag)}  &
  \colhead{stars}  \\
  \colhead{(1)} &
  \colhead{(2)}  &
  \colhead{(3)}  &
  \colhead{(4)}  &
  \colhead{(5)}  &
  \colhead{(6)}  &
  \colhead{(7)}  &
  \colhead{(8)}  &
  \colhead{(9)}
}
\startdata
      &      &                   &                  & 04:22-09:25 UT  &                 &                 &       &    \\
52171 & $u'$ & -20.952$\pm$0.020 & -0.011$\pm$0.010 & 0.475$\pm$0.015 &     \nodata     &     \nodata     & 0.018 & 20 \\
52171 & $g'$ & -22.640$\pm$0.013 & -0.006$\pm$0.012 & 0.184$\pm$0.008 &     \nodata     &     \nodata     & 0.009 & 19 \\
52171 & $r'$ & -22.605$\pm$0.009 & -0.137$\pm$0.016 & 0.093$\pm$0.005 &     \nodata     &     \nodata     & 0.007 & 27 \\
52171 & $i'$ & -22.153$\pm$0.011 & -0.150$\pm$0.025 & 0.061$\pm$0.005 &     \nodata     &     \nodata     & 0.006 & 19 \\
52171 & $z'$ & -21.279$\pm$0.027 & -0.107$\pm$0.062 & 0.050$\pm$0.013 &     \nodata     &     \nodata     & 0.014 & 18 \\
      &      &                   &                  &                 &                 &                 &       &    \\
      &      &                   &                  & 23:54-02:54 UT  & 02:54-05:54 UT  & 05:54-09:29 UT  &       &    \\
52172 & $u'$ & -20.944$\pm$0.025 & -0.013$\pm$0.011 & 0.449$\pm$0.018 & 0.465$\pm$0.019 & 0.466$\pm$0.020 & 0.020 & 18 \\
52172 & $g'$ & -22.618$\pm$0.013 &  0.049$\pm$0.010 & 0.182$\pm$0.009 & 0.195$\pm$0.010 & 0.183$\pm$0.010 & 0.010 & 19 \\
52172 & $r'$ & -22.582$\pm$0.012 &  0.011$\pm$0.020 & 0.112$\pm$0.008 & 0.114$\pm$0.009 & 0.107$\pm$0.009 & 0.009 & 19 \\
52172 & $i'$ & -22.107$\pm$0.015 &  0.007$\pm$0.032 & 0.060$\pm$0.009 & 0.069$\pm$0.010 & 0.057$\pm$0.010 & 0.010 & 19 \\
52172 & $z'$ & -21.229$\pm$0.021 &  0.018$\pm$0.045 & 0.034$\pm$0.013 & 0.053$\pm$0.014 & 0.036$\pm$0.014 & 0.015 & 19 \\
      &      &                   &                  &                 &                 &                 &       &    \\
      &      &                   &                  & 00:35-03:35 UT  & 03:35-09:05 UT  &     \nodata     &       &    \\
52309 & $u'$ & -20.930$\pm$0.022 & -0.030$\pm$0.008 & 0.480$\pm$0.016 & 0.503$\pm$0.014 &     \nodata     & 0.017 & 17 \\
52309 & $g'$ & -22.531$\pm$0.013 &  0.055$\pm$0.010 & 0.170$\pm$0.009 & 0.173$\pm$0.008 &     \nodata     & 0.010 & 20 \\
52309 & $r'$ & -22.513$\pm$0.008 & -0.062$\pm$0.013 & 0.091$\pm$0.006 & 0.092$\pm$0.005 &     \nodata     & 0.006 & 17 \\
52309 & $i'$ & -21.988$\pm$0.012 &  0.022$\pm$0.026 & 0.056$\pm$0.008 & 0.054$\pm$0.007 &     \nodata     & 0.009 & 20 \\
52309 & $z'$ & -21.080$\pm$0.015 &  0.136$\pm$0.033 & 0.059$\pm$0.010 & 0.058$\pm$0.009 &     \nodata     & 0.011 & 19 \\
      &      &                   &                  &                 &                 &                 &       &    \\
      &      &                   &                  & 00:44-03:44 UT  & 03:44-09:14 UT  &     \nodata     &       &    \\
52310 & $u'$ & -20.908$\pm$0.013 & -0.039$\pm$0.005 & 0.486$\pm$0.010 & 0.463$\pm$0.009 &     \nodata     & 0.010 & 11 \\
52310 & $g'$ & -22.484$\pm$0.008 &  0.064$\pm$0.005 & 0.162$\pm$0.004 & 0.150$\pm$0.005 &     \nodata     & 0.005 & 14 \\
52310 & $r'$ & -22.491$\pm$0.007 & -0.041$\pm$0.010 & 0.090$\pm$0.003 & 0.088$\pm$0.004 &     \nodata     & 0.004 & 13 \\
52310 & $i'$ & -21.958$\pm$0.014 &  0.037$\pm$0.028 & 0.051$\pm$0.007 & 0.041$\pm$0.007 &     \nodata     & 0.009 & 14 \\
52310 & $z'$ & -21.073$\pm$0.015 &  0.189$\pm$0.027 & 0.056$\pm$0.007 & 0.059$\pm$0.008 &     \nodata     & 0.009 & 14 \\
      &      &                   &                  &                 &                 &                 &       &    \\
      &      &                   &                  & 00:26-03:26 UT  & 03:26-09:22 UT  &     \nodata     &       &    \\
52553 & $u'$ & -20.757$\pm$0.008 & -0.018$\pm$0.004 & 0.455$\pm$0.006 & 0.457$\pm$0.006 &     \nodata     & 0.007 & 21 \\
52553 & $g'$ & -22.458$\pm$0.013 &  0.020$\pm$0.011 & 0.184$\pm$0.008 & 0.176$\pm$0.007 &     \nodata     & 0.011 & 26 \\
52553 & $r'$ & -22.407$\pm$0.011 & -0.018$\pm$0.020 & 0.087$\pm$0.006 & 0.082$\pm$0.006 &     \nodata     & 0.009 & 27 \\
52553 & $i'$ & -21.931$\pm$0.013 & -0.043$\pm$0.029 & 0.046$\pm$0.007 & 0.034$\pm$0.007 &     \nodata     & 0.011 & 27 \\
52553 & $z'$ & -21.097$\pm$0.012 & -0.036$\pm$0.028 & 0.055$\pm$0.007 & 0.040$\pm$0.006 &     \nodata     & 0.010 & 27 \\
      &      &                   &                  &                 &                 &                 &       &    \\
      &      &                   &                  & 23:49-02:49 UT  & 02:49-05:49 UT  & 05:49-09:25 UT  &       &    \\
52554 & $u'$ & -20.828$\pm$0.023 & -0.025$\pm$0.008 & 0.500$\pm$0.018 & 0.506$\pm$0.016 & 0.496$\pm$0.016 & 0.016 & 20 \\
52554 & $g'$ & -22.485$\pm$0.011 &  0.015$\pm$0.008 & 0.194$\pm$0.009 & 0.189$\pm$0.008 & 0.181$\pm$0.008 & 0.007 & 18 \\
52554 & $r'$ & -22.465$\pm$0.015 & -0.055$\pm$0.022 & 0.120$\pm$0.011 & 0.114$\pm$0.010 & 0.109$\pm$0.010 & 0.010 & 22 \\
52554 & $i'$ & -21.985$\pm$0.018 & -0.063$\pm$0.040 & 0.069$\pm$0.014 & 0.054$\pm$0.012 & 0.063$\pm$0.012 & 0.011 & 22 \\
52554 & $z'$ & -21.095$\pm$0.015 &  0.001$\pm$0.034 & 0.039$\pm$0.013 & 0.018$\pm$0.010 & 0.022$\pm$0.012 & 0.009 & 20 \\
      &      &                   &                  &                 &                 &                 &       &    \\
      &      &                   &                  & 23:46-02:46 UT  & 02:46-05:46 UT  & 05:46-08:50 UT  &       &    \\
52557 & $u'$ & -20.801$\pm$0.032 & -0.011$\pm$0.006 & 0.453$\pm$0.027 & 0.509$\pm$0.025 & 0.504$\pm$0.026 & 0.009 & 13 \\
52557 & $g'$ & -22.426$\pm$0.026 & -0.000$\pm$0.012 & 0.136$\pm$0.021 & 0.150$\pm$0.020 & 0.148$\pm$0.020 & 0.011 & 15 \\
52557 & $r'$ & -22.475$\pm$0.022 & -0.072$\pm$0.020 & 0.129$\pm$0.018 & 0.129$\pm$0.017 & 0.124$\pm$0.016 & 0.009 & 17 \\
52557 & $i'$ & -22.005$\pm$0.025 & -0.127$\pm$0.033 & 0.084$\pm$0.020 & 0.080$\pm$0.019 & 0.075$\pm$0.019 & 0.010 & 16 \\
52557 & $z'$ & -21.095$\pm$0.025 & -0.042$\pm$0.030 & 0.039$\pm$0.019 & 0.053$\pm$0.018 & 0.045$\pm$0.018 & 0.009 & 15 \\
      &      &                   &                  &                 &                 &                 &       &    \\
      &      &                   &                  & 23:56-02:56 UT  & 02:56-05:56 UT  & 05:56-09:25 UT  &       &    \\
52558 & $u'$ & -20.818$\pm$0.021 & -0.000$\pm$0.010 & 0.513$\pm$0.015 & 0.510$\pm$0.016 & 0.498$\pm$0.016 & 0.017 & 22 \\
52558 & $g'$ & -22.454$\pm$0.012 & -0.009$\pm$0.009 & 0.173$\pm$0.007 & 0.173$\pm$0.008 & 0.169$\pm$0.008 & 0.008 & 21 \\
52558 & $r'$ & -22.440$\pm$0.013 & -0.153$\pm$0.022 & 0.087$\pm$0.007 & 0.087$\pm$0.008 & 0.082$\pm$0.008 & 0.009 & 23 \\
52558 & $i'$ & -21.940$\pm$0.012 & -0.041$\pm$0.024 & 0.049$\pm$0.006 & 0.042$\pm$0.007 & 0.045$\pm$0.007 & 0.007 & 20 \\
52558 & $z'$ & -21.048$\pm$0.020 &  0.009$\pm$0.043 & 0.024$\pm$0.012 & 0.018$\pm$0.012 & 0.012$\pm$0.013 & 0.012 & 23 \\
      &      &                   &                  &                 &                 &                 &       &    \\
      &      &                   &                  &                 &                 &                 &       &    \\
\hline
      &      &                   &                  &                 &                 &                 &       &    \\
Ave.  & $u'$ & -20.829$\pm$0.006 & -0.021$\pm$0.002 & 0.472$\pm$0.003 &                 &                 &       &    \\
Ave.  & $g'$ & -22.508$\pm$0.004 &  0.035$\pm$0.003 & 0.172$\pm$0.002 &                 &                 &       &    \\
Ave.  & $r'$ & -22.507$\pm$0.004 & -0.062$\pm$0.006 & 0.093$\pm$0.001 &                 &                 &       &    \\
Ave.  & $i'$ & -22.014$\pm$0.005 & -0.044$\pm$0.010 & 0.052$\pm$0.002 &                 &                 &       &    \\
Ave.  & $z'$ & -21.105$\pm$0.006 &  0.041$\pm$0.012 & 0.044$\pm$0.002 &                 &                 &       &    \\

\enddata

\tablenotetext{*} {The second order extinction term values are
$-2.1 \times  10^{-2}$, $-1.6\times 10^{-2}$,  $-4.0 \times  10^{-3}$,
$6.0  \times 10^{-3}$   and    $3.0    \times  10^{-3}$,  for      the
$u'$,$g'$,$r'$,$i'$ and $z'$ respectively. These values are set to the
determined coefficients from \citet{Smith02}. }
\end{deluxetable}

\clearpage

\begin{deluxetable}{cc@{ }ccccccc@{ }c@{ }c@{ }c@{ }ccc@{ }c@{ }c@{ }c@{ }c}
\rotate
\tabletypesize{\tiny}
\tablecaption{The $u'g'r'i'z'$ Observations for CDFS Local Standard Stars:
   Calibrated Magnitudes and Colors. \label{stdstars}}
\tablewidth{0pt}
\tablenum{4}
\tablehead{
 \colhead{} &
  \colhead{}  &
  \colhead{}  &
  \colhead{}  &
  \colhead{}  &
  \colhead{}  &
  \colhead{}  &
  \colhead{}  &
    \multicolumn{5}{c}{Estimate of rms Mean Error}  &
  \colhead{}  &
   \multicolumn{5}{c}{ Number of Observations}  \\
\cline{9-13}\cline{15-19}\\
  \colhead{COMBO-17 \#} &
  \colhead{RA}  &
  \colhead{DEC}  &
  \colhead{$r'$}  &
  \colhead{$u'-g'$}  &
  \colhead{$g'-r'$}  &
  \colhead{$r'-i'$}  &
  \colhead{$i'-z'$}  &
  \colhead{$\sigma_{r'}$}  &
  \colhead{$\sigma_{u'-g'}$}  &
  \colhead{$\sigma_{g'-r'}$}  &
  \colhead{$\sigma_{r'-i'}$}  &
  \colhead{$\sigma_{i'-z'}$}  &
  \colhead{}  &
  \colhead{$n_{u'}$}  &
  \colhead{$n_{g'}$}  &
  \colhead{$n_{r'}$}  &
  \colhead{$n_{i'}$}  &
  \colhead{$n_{z'}$} \\
  \colhead{} &
  \colhead{(2000.0)}  &
  \colhead{(2000.0)}  &
  \colhead{}  &
  \colhead{}  &
  \colhead{}  &
  \colhead{}  &
  \colhead{}  &
  \colhead{}  &
  \colhead{}  &
  \colhead{}  &
  \colhead{}  &
  \colhead{}  &
  \colhead{}  &
  \colhead{}  &
  \colhead{}  &
  \colhead{}  &
  \colhead{}  &
  \colhead{}  \\
  \colhead{(1)} &
  \colhead{(2)}  &
  \colhead{(3)}  &
  \colhead{(4)}  &
  \colhead{(5)}  &
  \colhead{(6)}  &
  \colhead{(7)}  &
  \colhead{(8)}  &
  \colhead{(9)}  &
  \colhead{(10)}  &
  \colhead{(11)}  &
  \colhead{(12)}  &
  \colhead{(13)}  &
  \colhead{}  &
  \colhead{(14)}  &
  \colhead{(15)}  &
  \colhead{(16)}  &
  \colhead{(17)}  &
  \colhead{(18)}
}
\startdata
24094           &    03 32 24.69 &    -27 53 59.6 &      13.550 &    1.303 &    0.435 &    0.147 &    0.031 &       0.002 &    0.004 &    0.004 &    0.004 &    0.004 &   &       15 &    16 &    16 &    16 &    16 \\
34469           &    03 32 50.42 &    -27 48 33.1 &      13.838 &    1.319 &    0.435 &    0.140 &    0.046 &       0.001 &    0.005 &    0.003 &    0.003 &    0.004 &   &       16 &    16 &    16 &    16 &    16 \\
47831           &    03 32 40.22 &    -27 42 23.7 &      13.885 &    1.803 &    0.656 &    0.257 &    0.112 &       0.003 &    0.012 &    0.006 &    0.006 &    0.006 &   &       15 &    16 &    16 &    16 &    16 \\
23984           &    03 32 04.05 &    -27 53 54.9 &      14.640 &    1.172 &    0.402 &    0.137 &    0.007 &       0.005 &    0.015 &    0.007 &    0.007 &    0.008 &   &       15 &    16 &    16 &    16 &    16 \\
38206           &    03 32 52.34 &    -27 46 26.0 &      15.179 &    2.146 &    0.866 &    0.316 &    0.161 &       0.003 &    0.043 &    0.009 &    0.005 &    0.007 &   &       15 &    16 &    16 &    16 &    16 \\
27534           &    03 32 55.60 &    -27 51 26.2 &      15.248 &    1.497 &    0.560 &    0.198 &    0.084 &       0.003 &    0.021 &    0.006 &    0.004 &    0.009 &   &       14 &    14 &    14 &    14 &    14 \\
39797           &    03 32 10.44 &    -27 45 06.8 &      15.308 &    1.245 &    0.467 &    0.172 &    0.070 &       0.004 &    0.023 &    0.007 &    0.006 &    0.010 &   &       16 &    16 &    16 &    16 &    16 \\
46764           &    03 32 13.77 &    -27 42 13.6 &      15.736 &    1.673 &    0.818 &    0.358 &    0.180 &       0.009 &    0.169 &    0.016 &    0.013 &    0.013 &   &       11 &    16 &    16 &    16 &    16 \\
41920           &    03 32 08.13 &    -27 44 17.5 &      15.774 &    1.045 &    0.332 &    0.098 &    0.039 &       0.004 &    0.017 &    0.008 &    0.010 &    0.017 &   &       15 &    16 &    16 &    16 &    16 \\
27876           &    03 32 55.46 &    -27 51 06.4 &      15.873 &    2.074 &    1.104 &    0.490 &    0.239 &       0.007 &    0.089 &    0.016 &    0.010 &    0.009 &   &        7 &    14 &    14 &    14 &    14 \\
20195           &    03 32 49.63 &    -27 54 54.0 &      15.898 &    1.397 &    0.564 &    0.244 &    0.088 &       0.010 &    0.033 &    0.016 &    0.013 &    0.013 &   &       13 &    14 &    14 &    13 &    14 \\
26202           &    03 32 32.88 &    -27 51 47.8 &      16.375 &    1.006 &    0.361 &    0.117 &    0.057 &       0.009 &    0.037 &    0.013 &    0.014 &    0.018 &   &       13 &    14 &    14 &    14 &    13 \\
21659           &    03 32 54.69 &    -27 54 01.8 &      16.445 &    0.996 &    1.034 &    0.975 &    0.438 &       0.017 &    0.115 &    0.023 &    0.019 &    0.014 &   &        9 &    13 &    14 &    14 &    14 \\
44059           &    03 32 10.22 &    -27 43 06.9 &      16.696 &    2.142 &    0.802 &    0.362 &    0.109 &       0.015 &    0.023 &    0.025 &    0.021 &    0.035 &   &        3 &    16 &    16 &    16 &    14 \\
46429           &    03 32 31.60 &    -27 42 08.2 &      17.191 &    1.550 &    0.709 &    0.341 &    0.212 &       0.022 &    0.197 &    0.045 &    0.037 &    0.064 &   &        4 &    14 &    16 &    16 &    10 \\
43791           &    03 32 40.75 &    -27 43 18.4 &      17.329 &  \nodata &    0.815 &    0.345 &    0.188 &       0.019 &  \nodata &    0.041 &    0.023 &    0.058 &   &        0 &    11 &    15 &    12 &     7 \\
45812           &    03 32 33.19 &    -27 42 21.2 &      17.345 &  \nodata &    1.087 &    0.660 &    0.311 &       0.017 &  \nodata &    0.034 &    0.023 &    0.052 &   &        0 &    11 &    16 &    16 &    11 \\
38427           &    03 32 06.24 &    -27 45 42.4 &      17.424 &    0.866 &    0.428 &    0.219 &    0.183 &       0.019 &    0.225 &    0.031 &    0.033 &    0.089 &   &        9 &    16 &    16 &    11 &     7 \\
35104           &    03 32 53.42 &    -27 47 20.4 &      17.535 &  \nodata &    1.269 &    0.810 &    0.420 &       0.023 &  \nodata &    0.057 &    0.035 &    0.035 &   &        0 &     9 &    16 &    16 &    12 \\
30965           &    03 32 22.34 &    -27 49 25.3 &      17.759 &    1.112 &    0.992 &    0.380 &    0.254 &       0.020 &  \nodata &    0.026 &    0.028 &    0.047 &   &        1 &     8 &    13 &     8 &     6 \\
22486           &    03 32 10.87 &    -27 53 29.4 &      17.814 &  \nodata &    0.736 &    0.227 &    0.047 &       0.027 &  \nodata &    0.037 &    0.031 &  \nodata &   &        0 &     8 &    12 &     8 &     1 \\
21747           &    03 32 15.96 &    -27 53 49.9 &      17.980 &  \nodata &    1.155 &    0.520 &    0.321 &       0.026 &  \nodata &    0.043 &    0.032 &    0.105 &   &        0 &     7 &    10 &     8 &     6 \\
\enddata

\end{deluxetable}

\clearpage

\begin{deluxetable}{lr@{ }rr@{ }}
\tabletypesize{\footnotesize}
\tablecaption{Designation Cross Reference Table \label{crossref}}
\tablewidth{0pt}
\tablenum{5}
\tablehead{
  \colhead{COMBO-17 \#} &
  \colhead{RA (J2000.0)}  &
  \colhead{DEC (J2000.0)}  &
  \colhead{EIS Reference \#}
}
\startdata
24094           &    03 32 24.69 &    -27 53 59.6 &     J033224.70-275400.1  \\
34469           &    03 32 50.42 &    -27 48 33.1 &     J033250.45-274833.4  \\
47831           &    03 32 40.22 &    -27 42 23.7 &     J033240.26-274224.0  \\
23984           &    03 32 04.05 &    -27 53 54.9 &     J033204.08-275355.1  \\
38206           &    03 32 52.34 &    -27 46 26.0 &     J033252.36-274626.6  \\
27534           &    03 32 55.60 &    -27 51 26.2 &     J033255.63-275126.4  \\
39797           &    03 32 10.44 &    -27 45 06.8 &     J033210.46-274507.5  \\
46764           &    03 32 13.77 &    -27 42 13.6 &     J033213.80-274213.9  \\
41920           &    03 32 08.13 &    -27 44 17.5 &     J033208.16-274417.8  \\
27876           &    03 32 55.46 &    -27 51 06.4 &     J033255.51-275106.6  \\
20195           &    03 32 49.63 &    -27 54 54.0 &     J033249.66-275454.1  \\
26202           &    03 32 32.88 &    -27 51 47.8 &     J033232.88-275148.3  \\
21659           &    03 32 54.69 &    -27 54 01.8 &     J033254.70-275401.8  \\
44059           &    03 32 10.22 &    -27 43 06.9 &     J033210.27-274307.2  \\
46429           &    03 32 31.60 &    -27 42 08.2 &     J033231.64-274208.1  \\
43791           &    03 32 40.75 &    -27 43 18.4 &     J033240.76-274318.6  \\
45812           &    03 32 33.19 &    -27 42 21.2 &     J033233.21-274221.5  \\
38427           &    03 32 06.24 &    -27 45 42.4 &     J033206.26-274542.7  \\
35104           &    03 32 53.42 &    -27 47 20.4 &     J033253.43-274720.8  \\
30965           &    03 32 22.34 &    -27 49 25.3 &     J033222.35-274925.7  \\
22486           &    03 32 10.87 &    -27 53 29.4 &     J033210.88-275329.8  \\
21747           &    03 32 15.96 &    -27 53 49.9 &     J033216.00-275350.2  \\

\enddata

\end{deluxetable}

\clearpage

\clearpage

\clearpage

\clearpage

\clearpage

\clearpage

\clearpage

\clearpage

\clearpage

\clearpage

\end{document}